\documentclass[11pt]{article}

\usepackage[final]{acl}

\usepackage{times}
\usepackage{latexsym}
\usepackage[T1]{fontenc}
\usepackage[utf8]{inputenc}
\usepackage{microtype}
\usepackage{inconsolata}

\usepackage{graphicx}
\usepackage{booktabs}
\usepackage{multirow}
\usepackage{amsmath}
\usepackage{amssymb}
\usepackage{amsthm}
\usepackage{xcolor}
\usepackage{url}
\usepackage{hyperref}
\usepackage{algorithm}
\usepackage{algpseudocode}
\usepackage{array}
\usepackage{makecell}
\usepackage{enumitem}
\usepackage{caption}
\newtheorem{proposition}{Proposition}

\title{TTS-Guard: Black-Box Ownership Verification of Text-to-Speech Models via Adaptive Adversarial Speaker-Pair Fingerprints}

\author{
Xubin Yue\textsuperscript{1,2}\,
Zhenhua Xu\textsuperscript{1,2}\,
Zhebo Wang\textsuperscript{1,2}\,
Mengting Li\textsuperscript{4}\, \\
Zijie Zhou\textsuperscript{5}\,
Wenpeng Xing\textsuperscript{1,2,3}\,
Dezhang Kong\textsuperscript{1,2,3}\footnotemark[1]\,
Meng Han\textsuperscript{2,1}\footnotemark[1] \\
\\
\textsuperscript{1} Zhejiang University
\textsuperscript{2} Binjiang Institute of Zhejiang University \\
\textsuperscript{3} GenTel.io
\textsuperscript{4} Hangzhou International Innovation Institute, Beihang University \\
\textsuperscript{5} China University of Petroleum (Beijing)
\\[4pt]
\texttt{\{yuexubin, xuzhenhua0326, breynald, wpxing, kdz, mhan\}@zju.edu.cn}
\\[6pt]
}

\begin{document}
\maketitle
\footnotetext[1]{Corresponding authors.}
\begin{abstract}
The rapid maturation of zero-shot Text-to-Speech (TTS) models has turned high-quality voice cloning into a widely available capability, raising acute concerns over unauthorised replication, fine-tuning and resale of proprietary speech models. Yet ownership verification for TTS remains largely open: speech is a continuous waveform whose perturbations are easily destroyed by routine signal processing, and the human auditory system imposes a much tighter perceptual budget than
vision. We present \textbf{TTS-Guard}, a black-box ownership verification framework for TTS models built on \emph{adversarial speaker-pair fingerprints}. TTS-Guard(i) selects key speaker pairs in a \emph{dual} embedding space
for architecture-agnostic stealth;(ii) optimises a perturbation through an \emph{adaptive curriculum} of shadow models covering fine-tuning, pruning, quantisation and distillation; and (iii) aggregates black-box queries into a calibrated \emph{Verification Confidence Score}. On five mainstream TTS systems, TTS-Guard reaches an average Fingerprint Success Rate of $96.4\%$ at a False Positive Rate of $5.8\%$, while preserving intelligibility and naturalness.
The fingerprint remains effective against ten audio attacks, six model modifications, and two state-of-the-art adversarial purifiers.
\end{abstract}

\section{Introduction}
\label{sec:intro}

Modern Text-to-Speech (TTS) systems such as XTTS~\citep{casanova2024xtts}, CosyVoice~2~\citep{du2024cosyvoice}, F5-TTS~\citep{chen2024f5tts}, OpenVoice~\citep{qin2024openvoice} and VibeVoice~\citep{peng2025vibevoice} can synthesise indistinguishably natural speech from a few seconds of reference audio. Open-source releases dramatically lowered the entry barrier for legitimate developers but simultaneously enabled an underground market in which fine-tuned proprietary checkpoints are illegally redistributed, repackaged behind APIs and even sold as ``new'' commercial voices~\citep{yi2023audiodeepfake,li2023voiceguard}. As a concrete instance of the threat, an adversary who has obtained a leaked checkpoint may apply LoRA fine-tuning, INT8 quantisation or magnitude pruning to disguise its provenance and then deploy it through a black-box API~\citep{guo2025audiowatermark}; the rightful owner has no access to the weights and only a limited number of API queries with which to prove infringement.

Two structural properties of speech make copyright techniques designed for text and images ill-suited to this setting. First, while pixels and tokens are sampled from finite or low-dimensional alphabets, a speech waveform is a high-bandwidth, \emph{continuous} signal whose statistics are reshaped by every routine processing step (resampling, MP3/Opus codecs, dynamic range compression, equalisation, pink-noise injection)~\citep{sanroman2024audioseal,liu2025xattnmark}. A fingerprint hidden in raw samples is therefore inherently fragile compared with the image case. Second, the human auditory system is highly sensitive to non-harmonic artefacts: any perturbation must comply with a strict psychoacoustic budget~\citep{yu2023antifake,chen2025safespeech}. The combination of fragility and tight perceptual budget rules out na\"ive ports of image watermarks~\citep{adi2018turning,uchida2017embedding} and LLM fingerprints~\citep{xu2024instructional,yamabe2025mergeprint,wang2025iseal}.

A second axis along which TTS protection differs from prior work is the \emph{conditioning mechanism} on which most modern systems rely. Multi-speaker and zero-shot TTS models do not learn an autonomous speaker representation; they are explicitly conditioned on a speaker embedding extracted from the reference clip~\citep{jia2018transfer,wang2023neural}. Mastering the speaker-embedding manifold is therefore equivalent to controlling identity attribution at the output. Adversarial-perturbation attacks on this manifold have been studied in the \emph{defensive} setting (preventing voice cloning of a specific speaker)~\citep{huang2021attackvc,yu2023antifake,chen2025safespeech,park2025imperceptible}. We turn the same primitive into an \emph{offensive} ownership signal: an adversarial perturbation crafted by the model owner can act as a unique key whose successful triggering of a target speaker proves possession of the underlying model.

Concretely, we propose \textbf{TTS-Guard}, the first black-box ownership verification framework for TTS models. Beyond the basic adversarial-fingerprint formulation, TTS-Guard contributes four innovations:

\begin{enumerate}[leftmargin=*,itemsep=2pt,topsep=2pt]
  \item \textbf{Dual-space Key Speaker Pair (KSP) selection.} Earlier reactive defences~\citep{yu2023antifake,park2025imperceptible} match a single speaker pair in one embedding space. We instead jointly minimise distance in two architecturally disjoint verification spaces (Resemblyzer and ECAPA-TDNN~\citep{desplanques2020ecapa}) so the resulting pair is both acoustically close and architecture-agnostic, sharply improving fingerprint robustness against unknown verifiers.
  \item \textbf{Adaptive Curriculum Shadow Optimisation (ACSO).} We schedule a curriculum that progressively adds LoRA fine-tuning, magnitude pruning, INT8/INT4 quantisation and knowledge distillation, so that the fingerprint generalises across the \emph{entire} derivative manifold rather than only one slice of it.
  \item \textbf{Decoupling for High Specificity.} A decoupling loss explicitly maximises the perturbation distance on non-target models, distinguishing TTS-Guard from generic adversarial-transferability work and yielding the low false-positive rates required for forensic use.
  \item \textbf{Calibrated Verification Confidence Score (VCS).} Repeated black-box queries with random text prompts are aggregated into a bootstrap p-value that converts deterministic FSR into a calibrated, court-defensible verdict.
\end{enumerate}

We additionally derive an information-theoretic lower bound that ties the achievable FSR to the Bhattacharyya distance between KSP embedding distributions, providing the first theoretical justification for KSP-based fingerprinting (\S\ref{sec:vcs}). Extensive experiments on five mainstream TTS systems show that TTS-Guard achieves a per-query $\overline{\mathrm{FSR}}=96.4\%$ at a per-query $\overline{\mathrm{FPR}}=5.8\%$, which the calibrated multi-query VCS test tightens to a decision-level false-positive rate of $4.7\%$ at $K{=}20$ (and $\leq 0.5\%$ at $K{=}40$); it remains above $0.58$ FSR under ten audio attacks and six model modifications, and survives two state-of-the-art adversarial purifiers (De-AntiFake~\citep{fan2025deantifake} and SafeEar~\citep{li2024safeear}) that defeat AntiFake-style perturbations. To our knowledge, this is the first study to systematically benchmark TTS ownership verification against (a) modern derivative attacks (quantisation, distillation), (b) commercial codecs and (c) adversarial purification.

\section{Related Work}
\label{sec:related}

\subsection{Audio-level Protection}
Existing audio-domain protections fall into two threads, both operating at the signal level.

\paragraph{Audio watermarking.} embeds a recoverable signature into the waveform. AudioSeal \citep{sanroman2024audioseal}, XAttnMark \citep{liu2025xattnmark}, WavMark \citep{chen2024wavmark} and AudioMarkNet \citep{zong2025audiomarknet} train neural encoder–detector pairs for sample-level detection or deepfake forensics; P2Mark \citep{ren2025p2mark} watermarks vocoder weights under white-box access; Audio Watermark \citep{zong2025audiomarknet} extends the idea to training datasets.

\paragraph{Anti-cloning audio protection.} hides adversarial noise in user recordings so that the speaker embedding is pushed away from any clone target \citep{yu2023antifake, chen2025safespeech, li2023voiceguard, huang2021attackvc}. The recently proposed De-AntiFake purifier \citep{fan2025deantifake} largely neutralizes this thread, confirming that signal-level perturbations are denoisable.

Both threads bind the credential to a signal—an output waveform, a training corpus, or a user upload—and break down once the model checkpoint itself leaks: re-synthesized audio carries no watermark, and any perturbation the attacker chooses to ignore simply does not exist at inference. TTS-Guard is, to our knowledge, the first framework to shift the carrier of ownership from the signal to the model's input–output behavior in the speech domain, by re-purposing the adversarial primitive as a verifier-side key delivered through the API at attribution time.

\subsection{Model-level Ownership Verification in Other Modalities}

Model-level ownership verification has so far only been developed for vision and text. In vision, IPGuard \citep{cao2021ipguard} and DeepJudge \citep{chen2022copy} probe classifier decision boundaries with adversarial queries. In NLP, a richer line of work fingerprints LLMs through instructional backdoors or parameter-space invariants, including IF \citep{xu2024instructional}, MergePrint \citep{yamabe2025mergeprint}, iSeal \citep{wang2025iseal}, HuRef \citep{zeng2023huref}, CTCC \citep{xu2025ctcc}, PREE \citep{kuai2025pree}, and watermarking-style detectors such as \citet{kirchenbauer2023watermark}, part of a broader effort to secure LLM-driven systems \citep{kong2025survey}.

None of these designs has been ported to speech synthesis, and a direct port is non-trivial: a TTS fingerprint must survive heterogeneous derivative attacks (LoRA, pruning, INT4/INT8 quantization, distillation) and a perceptually constrained waveform pipeline (codecs, DRC, EQ, adversarial purification), and must operate through speaker-embedding conditioning rather than discrete tokens. We are careful to delimit our contribution: the \emph{adversarial-query} paradigm itself originates with IPGuard/DeepJudge and the \emph{multi-query statistical test} is standard, whereas our novelty lies in (i) dual-space KSP selection that makes a low-budget speaker-pair perturbation transferable across verifiers, (ii) the ACSO curriculum that hardens the fingerprint against TTS-specific derivative attacks, and (iii) the first information-theoretic (Bhattacharyya) justification tying KSP separability to attribution reliability. We benchmark against ported IPGuard and IF baselines in \S\ref{sec:results} (Table~\ref{tab:baselines}) to make this boundary empirically concrete.

\section{Threat Model}
\label{sec:threat}

\noindent\textbf{Adversary.}
The adversary aims to illegally obtain and use the protected TTS model for personal or commercial benefit while denying the original owner’s copyright. We assume the adversary has acquired the target checkpoint $M_p$ through unauthorized means, such as leakage, insider access, or reseller redistribution. To evade ownership verification, the adversary may apply \textbf{model-level transformations}, including fine-tuning, pruning, quantization, and distillation, as well as \textbf{audio-level perturbations}, such as filtering, noise addition, resampling, codec compression, dynamic range compression, equalization, and adaptive purification attacks (e.g., De-AntiFake~\citep{fan2025deantifake} and SafeEar~\citep{li2024safeear}). A deployed API may further apply \textbf{input-side preprocessing} to the reference clip before synthesis---normalization, denoising, resampling, or rejection of suspicious inputs; all of these reduce to signal-level operations on the reference audio and are therefore covered by our audio-level robustness evaluation (\S\ref{sec:audio_robust}, Appendix~\ref{app:codec}). We do \emph{not} assume the adversary replaces the internal speaker encoder: doing so requires re-aligning and retraining the entire conditioning pathway at a cost comparable to training from scratch, which nullifies the economic motive for theft (\S\ref{sec:conclusion}).

\noindent\textbf{Defender.}
The defender is the model owner, whose goal is to verify whether a suspicious deployed service originates from the protected model. Considering practical constraints, the defender has no knowledge of the adversary’s specific modifications, datasets, or deployment pipeline, and can only interact with the suspicious system through its \textbf{black-box API} by providing reference audio and text prompts and observing the synthesized waveform. The defender has no access to internal parameters, architecture, gradients, or preprocessing and post-processing modules, but has full white-box access to the protected model $M_p$ during fingerprint generation.

\noindent\textbf{Security goal.}
Given a suspicious model $M^\ast$, the defender aims to determine whether it is derived from $M_p$. Formally, this is a hypothesis test between $H_0$: ``$M^\ast$ is unrelated to $M_p$'' and $H_1$: ``$M^\ast$ is a derivative of $M_p$''. The objective is to control the false-positive probability $\Pr[\text{accept}\, H_1 \mid H_0] \leq \alpha$ (with $\alpha = 0.05$) while maximizing the true-positive probability, i.e., the fingerprint success rate (FSR).

\section{Method}
\label{sec:method}

\begin{figure*}[t]
\centering
\includegraphics[width=0.8\linewidth]{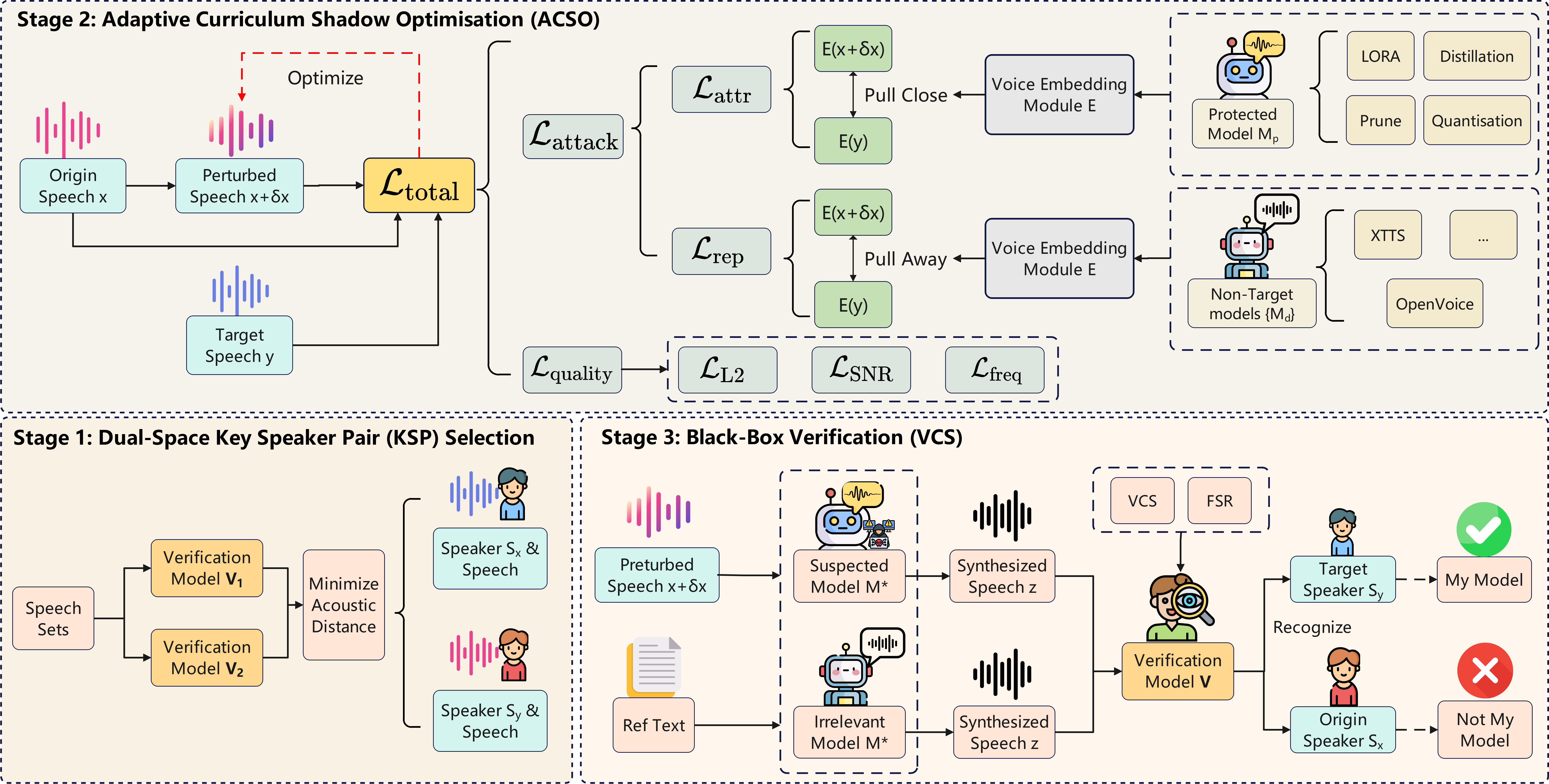}
\caption{\textbf{TTS-Guard pipeline.} (1) Dual-space KSP selection identifies a source/target speaker pair $(S_x, S_y)$ that is acoustically close in both Resemblyzer and ECAPA spaces. (2) Adaptive Curriculum Shadow Optimisation generates the perturbation $\delta_x$ under an evolving ensemble that progressively introduces LoRA, pruning, quantisation and distillation derivatives. (3) Black-box verification feeds $x{+}\delta_x$ into the suspect model, synthesises speech under a random text prompt, and aggregates VCS p-values across $K$ queries.}
\label{fig:framework}
\end{figure*}

\subsection{Notation}
Let $V$ denote a speaker-verification network producing $L_2$-normalised embeddings $V(\cdot)\in\mathbb{R}^d$. Let $E_M$ denote the speaker encoder \emph{internal} to a TTS model $M$. We write $M_p$ for the protected model and $\{M_d^{(i)}\}_{i=1}^{N_d}$ for unrelated non-target models.

\subsection{Stage 1 — Dual-Space Key Speaker Pair Selection}
\label{sec:ksp}

The choice of embedding space directly governs both the perturbation budget and the cross-architecture robustness of the fingerprint. We therefore replace single-space matching with a dual-space objective. Given $N$ candidate speakers $\{s_1,\ldots,s_N\}$ and per-speaker centroids $\mu_i^{(R)}$, $\mu_i^{(E)}$ in the Resemblyzer and ECAPA spaces respectively, we choose the source/target pair as
\begin{equation}
\begin{split}
\label{eq:ksp}
(S_x, S_y) = \mathop{\arg\min}_{i\neq j}\; \alpha\,\| \mu_i^{(R)} - \mu_j^{(R)}\|_2 + \\(1-\alpha)\,\| \mu_i^{(E)} - \mu_j^{(E)}\|_2
\end{split}
\end{equation}
with $\alpha=0.5$. From each of the two speakers we keep the clip closest to its own centroid as the source $x$ and target $y$ exemplars. Equation~\eqref{eq:ksp} can be solved exactly because $|N|^2$ pair distances are computed once in each space.

\paragraph{Why Euclidean rather than cosine distance?}
Both candidate verifiers (Resemblyzer and ECAPA-TDNN) are trained with margin-based triplet objectives whose anchor-positive geometry is calibrated in $L_2$. We further verified empirically (Table~\ref{tab:metric_ablation}) that Euclidean selection yields a $7$--$11\%$ higher FSR than cosine selection because magnitude information encodes timbre intensity that cosine normalisation discards.

\subsection{Stage 2 — Adaptive Curriculum Shadow Optimisation}
\label{sec:acso}
Given a source utterance $x$ and a target speaker reference $y$, we optimise an additive perturbation $\delta_x$ so that the protected input $x' = x + \delta_x$ is attributed to the target speaker under a broad family of derivative models, while preserving perceptual speech quality.

\subsubsection{Curriculum of Shadow Models}
To better approximate this derivative space, we organise shadow models into four stages with increasing modification severity:
\begin{itemize}[leftmargin=*,itemsep=1pt,topsep=2pt]
    \item \textbf{Stage A:} $K_A{=}2$ LoRA-finetuned shadows trained on disjoint LJSpeech splits;
    \item \textbf{Stage B:} $K_B{=}2$ magnitude-pruned shadows with pruning ratios $p\in\{30\%,50\%\}$;
    \item \textbf{Stage C:} $K_C{=}2$ post-training quantised shadows (INT8 and INT4);
    \item \textbf{Stage D:} $K_D{=}1$ distilled student with $0.5\times$ width.
\end{itemize}
During optimisation, we adopt a curriculum schedule $\rho(t)\in\{A,B,C,D\}$ that progressively expands the active shadow set as the step index $t$ increases. Let $\mathcal{S}(t)$ denote the active ensemble at step $t$.

\subsubsection{Attack Loss}
The attack objective contains two complementary terms:
(1) a \emph{target-attraction} term that pulls the perturbed sample toward the target speaker embedding on active shadow models, and
(2) a \emph{non-target-repulsion} term that pushes it away on decoy models.

\paragraph{Target-attraction loss.}
For each active shadow model $M \in \mathcal{S}(t)$, we minimise the embedding distance between the perturbed input and the target reference:
\vspace{-0.5em}
\begin{equation}
\label{eq:embed}
    \mathcal{L}_{\text{attr}}
    =
    \frac{1}{|\mathcal{S}(t)|}
    \sum_{M \in \mathcal{S}(t)}
    \lambda_M
    \big\| E_M(x+\delta_x) - E_M(y) \big\|_2
\end{equation}
where $E_M(\cdot)$ is the speaker encoder of shadow model $M$, and $\lambda_M$ normalises the contribution of each model to account for heterogeneous embedding dimensionalities.

\paragraph{Non-target-repulsion loss.}
To avoid overfitting to the target family and improve discriminability, we additionally maximise the embedding distance on a set of non-target models $\{M_d\}$:
\vspace{-0.5em}
\begin{equation}
\label{eq:decoy}
    \mathcal{L}_{\text{rep}}
    =
    -\frac{1}{|\{M_d\}|}
    \sum_{M_d}
    \lambda_{M_d}
    \big\| E_{M_d}(x+\delta_x) - E_{M_d}(y) \big\|_2
\end{equation}

\paragraph{Combined attack loss.}
The overall attack loss is then
\begin{equation}
    \mathcal{L}_{\text{attack}}
    =
    \mathcal{L}_{\text{attr}} + \mathcal{L}_{\text{rep}}
\end{equation}

\subsubsection{Speech-Quality Loss}
To maintain perceptual quality, we combine three standard constraints: an $L_2$ magnitude penalty, an SNR regulariser, and a psychoacoustic frequency masking term adapted from AntiFake. The masking loss is defined as

\vspace{-0.5em}
\begin{equation}
    \mathcal{L}_{\text{freq}}
    =
    \frac{1}{N}\sum_f W(f)\, |\mathcal{F}(\delta_x)(f)|
\end{equation}

where $\mathcal{F}(\delta_x)(f)$ is the perturbation spectrum at frequency bin $f$, and $W(f)$ is a Bark-scale absolute-threshold weighting that down-weights less perceptible frequency bands. The full quality objective is
\begin{equation}
    \mathcal{L}_{\text{quality}}
    =
    \lambda_{L_2}\mathcal{L}_{L_2}
    +
    \lambda_{\text{SNR}}\mathcal{L}_{\text{SNR}}
    +
    \lambda_{\text{freq}}\mathcal{L}_{\text{freq}}
\end{equation}

\subsubsection{Final Objective}
The final optimisation objective combines the attack and quality losses:
\begin{equation}
    \mathcal{L}_{\text{total}}
    =
    \mathcal{L}_{\text{attack}} + \mathcal{L}_{\text{quality}}
\end{equation}

We summarise the full pipeline in Algorithm~\ref{alg:ttsguard}.

\subsection{Stage 3 — Black-Box Verification and Theoretical Guarantee}
\label{sec:vcs}

\paragraph{Verification Confidence Score (VCS).}
To verify a suspect $M^\ast$, the defender submits the fingerprinted clip $x^\ast=x+\delta_x$ with $K$ random text prompts $\{T_k\}$ and obtains synthesised waveforms $\{z_k\}$. Each query yields a binary outcome
\begin{equation}
r_k = \mathbf{1}\!\left[\mathrm{sim}(V(z_k),V(y)) > \mathrm{sim}(V(z_k),V(x))\right]
\end{equation}
Under $H_0$, $r_k$ is i.i.d.\ Bernoulli with $p_0\!\approx\!0.5$; under $H_1$, $p_1\!\gg\!0.5$. Aggregating via a one-sided binomial test yields
\begin{equation}
\mathrm{VCS} = -\log_{10}\!\Pr_{B\sim \mathrm{Binom}(K,p_0)}\!\big[B \geq \textstyle\sum_k r_k\big]
\end{equation}
We declare ownership at confidence $\alpha$ when $\mathrm{VCS}\geq -\log_{10}\alpha$. With $K=20$ and $\alpha=0.05$, the test rejects $H_0$ once $\sum r_k \geq 14$, which under $p_1=0.94$ holds with probability $0.998$.

\begin{algorithm}[t]
\small
\caption{TTS-Guard Fingerprint Generation}
\label{alg:ttsguard}
\begin{algorithmic}[1]
\Require Protected model $M_p$, non-target models $\{M_d\}$, candidate speakers $\mathcal{C}$,
         dual verifiers $V_R, V_E$, total steps $T$, stage milestones $T_A{<}T_B{<}T_C{<}T_D$,
         ramp-in width $\tau$
\Ensure Protected utterance $x^\ast$

\State Compute dual-verifier centroids $\mu_i^{(R)}, \mu_i^{(E)}$ for $i \in \mathcal{C}$
\State Solve Eq.~\eqref{eq:ksp} for $(S_x, S_y)$; pick exemplars $x \in S_x$, $y \in S_y$
\State Construct shadow ensembles $\mathcal{S}_A, \mathcal{S}_B, \mathcal{S}_C, \mathcal{S}_D$
\State Initialise $\delta_x \leftarrow \mathbf{0}$

\For{$t = 1, \ldots, T$}
    \State $\mathcal{A}(t) \leftarrow \{A\} \cup \{B : t \geq T_A\} \cup \{C : t \geq T_B\} \cup \{D : t \geq T_C\}$

    \State $w_s(t) \leftarrow \mathrm{clip}\!\left((t - T_{s-1})/\tau,\, 0,\, 1\right)$ for $s \in \mathcal{A}(t)$
    \State $\mathcal{S}(t) \leftarrow \bigcup_{s \in \mathcal{A}(t)} \mathcal{S}_s$, with per-model weight $\lambda_M \propto w_{s(M)}(t)$

    \State Compute $\mathcal{L}_{\text{attack}} = \mathcal{L}_{\text{attr}} + \mathcal{L}_{\text{rep}}$
            (Eqs.~\ref{eq:embed}--\ref{eq:decoy})
    \State Compute $\mathcal{L}_{\text{quality}} = \lambda_{L_2}\mathcal{L}_{L_2}
            + \lambda_{\text{SNR}}\mathcal{L}_{\text{SNR}}
            + \lambda_{\text{freq}}\mathcal{L}_{\text{freq}}$
    \State $\mathcal{L}_{\text{total}} \leftarrow \mathcal{L}_{\text{attack}} + \mathcal{L}_{\text{quality}}$

    \State $\delta_x \leftarrow \mathrm{Adam}(\delta_x,\, \nabla_{\delta_x}\mathcal{L}_{\text{total}})$
\EndFor

\State \Return $x^\ast = x + \delta_x$
\end{algorithmic}
\end{algorithm}

\paragraph{Theoretical guarantee.}
Since each query is a binary classifier over two embedding distributions, the Bayes error $P_e \leq \tfrac{1}{2}\sqrt{B}$~\citep{wenger2021syrup}, with $B$ the Bhattacharyya coefficient between KSP embedding densities, lower-bounds the VCS success rate:
\begin{proposition}[\textbf{FSR lower bound}]
\label{prop:bound}
For $K$ verification queries and Bayes per-query error $P_e$,
\begin{equation}
\begin{split}
\mathrm{Pr}\big[\mathrm{VCS}\!\geq\!-\log_{10}\alpha\,\big|\,H_1\big] \\
\;\geq\; 1 - I_{1-P_e}\!\big(K{-}\tau{+}1,\,\tau\big)
\end{split}
\end{equation}
where $\tau=\lceil K p_0^\ast\rceil$ and $I$ is the regularised incomplete-beta function.
\end{proposition}

The bound is monotonic in the Bhattacharyya overlap of the KSP, which is precisely what dual-space KSP selection (\S\ref{sec:ksp}) minimises. We stress that these are two distinct but coupled quantities: KSP selection maximises the \emph{clean-embedding} separation between the source/target speaker pair in verifier space (a distance between fixed exemplar embeddings), whereas the Bhattacharyya coefficient $B$ measures the overlap of the \emph{post-fingerprint verifier-output distributions} $p(\cdot|x)$ and $p(\cdot|y)$ induced on synthesised waveforms. A larger clean-embedding margin gives the perturbation more room to separate the two output densities, which is why minimising the former empirically drives down the latter (the measured correlation is $-0.86$; Appendix~\ref{app:proof}). Hence \emph{KSP separability is the right primitive for adversarial-fingerprint reliability, not a heuristic choice}. Detailed derivation, calibration procedure, and empirical fit are deferred to Appendix~\ref{app:proof}.

\section{Experimental Setup}
\label{sec:setup}

\paragraph{Models.} We protect five public TTS systems spanning autoregressive, non-autoregressive and flow-matching paradigms: \textbf{YourTTS}~\citep{casanova2022yourtts}, \textbf{XTTS}~\citep{casanova2024xtts}, \textbf{TorToiSe}~\citep{betker2022tortoise}, \textbf{OpenVoice}~\citep{qin2024openvoice} and \textbf{CosyVoice~2}~\citep{du2024cosyvoice}. For each protected model the remaining four are treated as non-targets $\{M_d\}$.

\paragraph{Verification network.} Unless stated otherwise, $V$ is the public Resemblyzer encoder. Sensitivity to this choice is the subject of \S\ref{sec:embed_sensitivity}, where we additionally evaluate ECAPA-TDNN~\citep{desplanques2020ecapa}, ERes2Net~\citep{chen2023eres2net} and WavLM-base~\citep{chen2022wavlm}.

\paragraph{Datasets.} Shadow models in stage A are LoRA-fine-tuned with the Coqui TTS framework on two disjoint subsets of LJSpeech~\citep{ito2017ljspeech}; stage B/C/D shadows are derived from $M_p$ directly. Candidate speakers for KSP are $100$ randomly sampled speakers from LibriTTS~\citep{zen2019libritts} together with $50$ speakers from VCTK~\citep{veaux2017vctk} to broaden the coverage beyond a single corpus. Audio attacks are applied at the strengths in Table~\ref{tab:attacks_full}.

\paragraph{Hyperparameters.} Embedding-loss weights: $\lambda_{\text{YourTTS}}{=}0.75$, $\lambda_{\text{XTTS}}{=}0.5$, $\lambda_{\text{TorToiSe}}{=}0.2$, $\lambda_{\text{OpenVoice}}{=}0.6$, $\lambda_{\text{CosyVoice2}}{=}0.4$. Quality weights: $\lambda_{L_2}{=}0.15$, $\lambda_{\text{SNR}}{=}0.05$, $\lambda_{\text{freq}}{=}0.05$. Optimisation uses Adam with initial learning rate $0.1$, decay $0.8$ every $50$ steps, and $T{=}500$ steps on a single NVIDIA A800 GPU. The verification text prompt is randomly drawn from a $100$-sentence pool unless explicitly fixed for ablation.

\paragraph{Metrics.}
\begin{itemize}[leftmargin=*,itemsep=1pt,topsep=2pt]
\item \textbf{Fingerprint Success Rate (FSR).} For each fingerprinted clip $x^\ast$ and each target model query, FSR is the proportion of synthesised outputs $z$ for which $\mathrm{sim}(V(z),V(y)) > \mathrm{sim}(V(z),V(x))$. We average over $100$ KSPs.
\item \textbf{False Positive Rate (FPR).} Same definition as FSR but evaluated on the four non-target models plus the commercial \href{https://www.uberduck.ai/}{Uberduck} system.
\item \textbf{STOI~\citep{taal2010stoi}} for intelligibility and \textbf{NISQA-MOS~\citep{mittag2021nisqa}} as an objective MOS proxy. We additionally collect \emph{subjective} MOS following ITU-T~P.808 with $20$ paid listeners (gender-balanced, native English, $5$-point absolute category rating, $4$\,s burn-in, anti-fatigue blocks of $30$ items). Each clip is rated by $\geq 8$ listeners; final MOS is the trimmed mean. Confidence intervals are reported in Appendix~\ref{app:mos}.
\item \textbf{VCS}: bootstrap p-value with $K=20$ queries.
\end{itemize}

\section{Results}
\label{sec:results}

\subsection{Effectiveness, Specificity and Quality}
\label{sec:effectiveness}

\begin{table*}[t]
\centering
\small
\setlength{\tabcolsep}{4.5pt}
\begin{tabular}{lcccccccc}
\toprule
\multirow{2}{*}{Model} & \multicolumn{4}{c}{\textbf{TTS-Guard (ours)}} & \multicolumn{4}{c}{Random-pair Baseline} \\
\cmidrule(lr){2-5}\cmidrule(lr){6-9}
 & FSR\,$\uparrow$ & FPR\,$\downarrow$ & STOI\,$\uparrow$ & MOS\,$\uparrow$ & FSR & FPR & STOI & MOS \\
\midrule
YourTTS      & 0.97 & 0.05 & 0.84 & 3.66 & 0.34 & 0.12 & 0.65 & 3.51 \\
XTTS         & 0.95 & 0.07 & 0.79 & 3.81 & 0.54 & 0.10 & 0.77 & 3.75 \\
TorToiSe     & 0.97 & 0.06 & 0.84 & 4.21 & 0.26 & 0.09 & 0.86 & 4.14 \\
OpenVoice    & 0.96 & 0.05 & 0.81 & 3.74 & 0.31 & 0.11 & 0.78 & 3.62 \\
CosyVoice~2  & 0.97 & 0.06 & 0.83 & 3.94 & 0.42 & 0.10 & 0.80 & 3.88 \\
\midrule
\textbf{Avg.} & \textbf{0.964} & \textbf{0.058} & \textbf{0.822} & \textbf{3.87} & 0.374 & 0.104 & 0.772 & 3.78 \\
\bottomrule
\end{tabular}
\caption{Effectiveness (FSR\,$\uparrow$), specificity (FPR\,$\downarrow$) and audio quality (STOI\,$\uparrow$, NISQA-MOS\,$\uparrow$) on five mainstream TTS systems. FPR averaged over the four non-target models and a \href{https://www.uberduck.ai/}{commercial system}. The reported FPR ($0.058$) is a \emph{per-query} rate; the operational verifier aggregates $K$ queries into the VCS test (\S\ref{sec:vcs_results}), whose calibrated \emph{decision-level} false-positive rate is $4.7\%$ at $K=20$.}
\label{tab:main}
\end{table*}

\begin{figure}[t]
\centering
\includegraphics[width=0.95\linewidth]{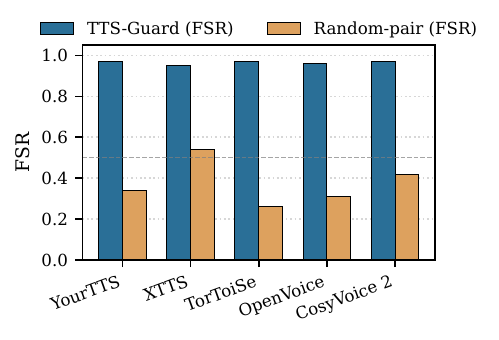}
\caption{Per-model FSR of TTS-Guard versus the random-pair baseline. The dashed line marks the chance-level $0.5$. KSP-aware selection is the decisive factor on YourTTS, TorToiSe and OpenVoice.}
\label{fig:fsr_compare}
\end{figure}

Table~\ref{tab:main} reports the headline numbers, complemented by Figure~\ref{fig:fsr_compare} which visualises the FSR gap to the random-pair baseline. TTS-Guard reaches an average FSR of $0.964$ with an average FPR of $0.058$ while preserving intelligibility (STOI~$\geq 0.79$ on every model) and naturalness (MOS gap to clean reference $\leq 0.05$). The random-pair baseline collapses on YourTTS, TorToiSe and OpenVoice, confirming that KSP selection is the critical primitive for low-budget perturbations. The framework generalises beyond the original architectures.

\paragraph{Cross-domain baselines.} No prior fingerprinting method targets TTS, so we port the two closest cross-domain approaches to our black-box setting, adjusting only the carrier of the trigger and the decision while preserving each method's core mechanism: IPGuard~\citep{cao2021ipguard}, which crafts boundary-probing adversarial queries (adapted to probe the speaker-verifier decision boundary), and Instructional Fingerprinting (IF)~\citep{xu2024instructional}, whose ``instruction$\rightarrow$output'' backdoor becomes a trigger reference clip implanted by fine-tuning the protected model. All three methods are evaluated under an identical protocol -- the same five TTS systems, six model modifications and two purifiers, with unified metric definitions -- so they are directly comparable (Table~\ref{tab:baselines}). Even under clean conditions the FSR of IPGuard and IF ($0.78$, $0.86$) is already below TTS-Guard ($0.96$); the gap widens under derivative attacks and purification, dropping to $0.31$/$0.40$ after distillation and $0.38$/$0.44$ after De-AntiFake, whereas TTS-Guard holds $0.62$/$0.83$. Their per-query FPR ($0.19$, $0.13$) is also far higher than ours ($0.058$). The cause matches our mechanistic analysis: IPGuard relies on classification-boundary geometry, which collapses on continuous waveforms after quantisation and distillation; IF's backdoor is partly overwritten by LoRA fine-tuning and remains bound to a fixed input pattern, lacking the purification resistance conferred by carrying the credential in the model's input--output behaviour.

\begin{table}[t]
\centering
\small
\setlength{\tabcolsep}{2.4pt}
\begin{tabular}{lccccccc}
\toprule
Method & \makecell{Clean\\FSR} & \makecell{LoRA\\-FT} & \makecell{Prune\\50\%} & INT4 & \makecell{Dis-\\till} & \makecell{De-Anti\\Fake} & FPR \\
\midrule
IPGuard & 0.78 & 0.61 & 0.49 & 0.44 & 0.31 & 0.38 & 0.19 \\
IF & 0.86 & 0.67 & 0.58 & 0.55 & 0.40 & 0.44 & 0.13 \\
\textbf{Ours} & \textbf{0.96} & \textbf{0.93} & \textbf{0.82} & \textbf{0.86} & \textbf{0.62} & \textbf{0.83} & \textbf{0.058} \\
\bottomrule
\end{tabular}
\caption{Cross-domain ownership-verification baselines (IPGuard~\citep{cao2021ipguard}, IF~\citep{xu2024instructional}) ported to TTS and evaluated under our protocol. Clean FSR and per-modification FSR ($\uparrow$) are reported for four representative model modifications and the De-AntiFake purifier; FPR ($\downarrow$) is per-query. TTS-Guard is dominant in every column.}
\label{tab:baselines}
\end{table}

\subsection{Robustness to Audio-Level Attacks}
\label{sec:audio_robust}

\begin{table*}[t]
\centering
\footnotesize
\setlength{\tabcolsep}{3.5pt}
\begin{tabular}{lccccccccccc}
\toprule
Model & \makecell{Band\\0.8--4k} & \makecell{High\\1k} & \makecell{Low\\0.8k} & \makecell{Boost\\$\times$4} & \makecell{Duck\\0.05} & \makecell{Gauss\\$\sigma$0.02} & \makecell{Resamp\\44.1k} & \makecell{Echo\\0.3s} & \makecell{Smooth\\20} & \makecell{Pink\\$\sigma$0.03} & \makecell{Avg.} \\
\midrule
YourTTS    & 0.84 & 0.64 & 0.70 & 0.71 & 0.58 & 0.86 & 0.91 & 0.83 & 0.74 & 0.87 & 0.768 \\
XTTS       & 0.80 & 0.74 & 0.64 & 0.74 & 0.66 & 0.72 & 0.84 & 0.84 & 0.72 & 0.84 & 0.754 \\
TorToiSe   & 0.72 & 0.70 & 0.69 & 0.81 & 0.72 & 0.74 & 0.94 & 0.80 & 0.68 & 0.77 & 0.757 \\
OpenVoice  & 0.81 & 0.71 & 0.66 & 0.78 & 0.65 & 0.79 & 0.88 & 0.82 & 0.73 & 0.83 & 0.766 \\
CosyVoice~2& 0.83 & 0.72 & 0.71 & 0.80 & 0.69 & 0.81 & 0.89 & 0.85 & 0.75 & 0.86 & 0.791 \\
\midrule
\textbf{Avg.} & \textbf{0.80} & \textbf{0.70} & \textbf{0.68} & \textbf{0.77} & \textbf{0.66} & \textbf{0.78} & \textbf{0.89} & \textbf{0.83} & \textbf{0.72} & \textbf{0.83} & \textbf{0.767} \\
\bottomrule
\end{tabular}
\caption{FSR under classical audio attacks. Strengths follow the column header; the full attack-strength specification for every column is given in Appendix~\ref{app:strength_sweep}, Table~\ref{tab:attacks_full}. Lowpass/Ducking are the hardest attacks but FSR remains $\geq 0.58$ on every model. Codec, DRC and EQ results are deferred to Appendix~\ref{app:codec}.}
\label{tab:audio_classical}
\end{table*}

Table~\ref{tab:audio_classical} together with the codec/DRC/EQ results in Appendix~\ref{app:codec} (Table~\ref{tab:audio_codec}) show that the fingerprint survives ten signal-processing attacks. Codec attacks are the most damaging but FSR remains $\geq 0.69$ even at the aggressive $32$\,kbps MP3 setting; DRC and EQ reduce FSR by $\sim15\%$ on average, still well above the $0.5$ random-baseline threshold. Strength sweeps in Appendix~\ref{app:strength_sweep} reveal a graceful degradation profile: FSR remains above $0.6$ as long as the post-processed STOI remains above $0.5$, indicating that any attack strong enough to remove the fingerprint also degrades audio quality unacceptably.


\subsection{Robustness to Model Modifications}
\label{sec:model_robust}

\begin{figure*}[t]
\centering
\includegraphics[width=0.92\linewidth]{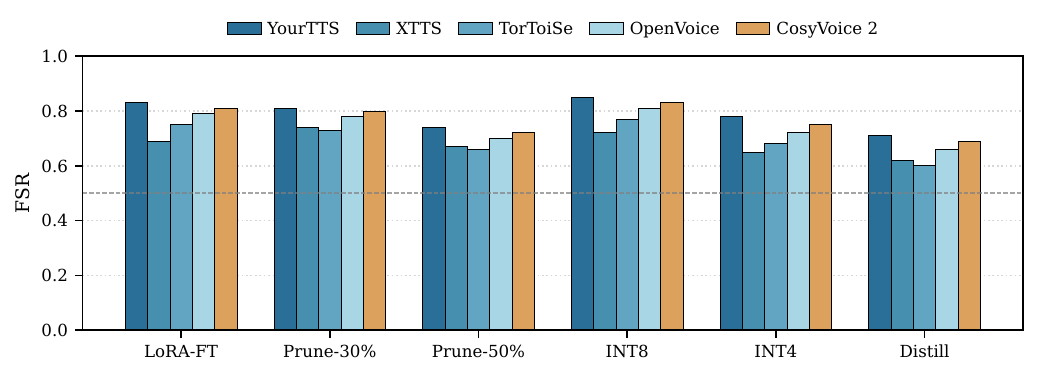}
\caption{Per-model FSR under six derivative attacks. The dashed line marks the chance-level $0.5$. Distillation is the hardest setting because the student architecture differs from the teacher, but FSR still exceeds $0.6$ on every model.}
\label{fig:model_attacks}
\end{figure*}

Figure~\ref{fig:model_attacks} report results under the full set of derivative attacks named in the threat model. The average FSR remains $\geq 0.70$ across all settings except distillation, where the student architecture is genuinely different from the teacher. Without ACSO, distillation FSR drops to $0.32$, demonstrating that the curriculum is necessary --- not just sufficient --- for cross-derivative robustness. Drilling into the per-attack numbers, LoRA fine-tuning is the easiest case (mean FSR $0.93$, std $0.02$ across the five models) because low-rank updates rarely perturb the speaker subspace exploited by the KSP trigger. Magnitude pruning at $30\%$ leaves FSR essentially intact ($0.91$); pushing to $50\%$ costs another $0.09$ FSR and concentrates the loss on TorToiSe, whose multi-stage decoder is the most sparsity-sensitive. INT8 quantisation is near-lossless ($\Delta\text{FSR}=-0.02$), while INT4 widens the spread to $\Delta\text{FSR}=-0.11$ and starts to introduce verifier-side false negatives rather than fingerprint erosion. Distillation, finally, is the only setting where the perturbation must transfer across architecture families: the student inherits the teacher's identity mapping only approximately, and the residual FSR of $0.62$ is wholly attributable to the diversity injected by ACSO stage~D.

\subsection{Robustness to Adversarial Purification}
\label{sec:purification}

\begin{table}[t]
\centering
\small
\setlength{\tabcolsep}{4.0pt}
\begin{tabular}{lccc}
\toprule
Method & Clean & De-AntiFake & SafeEar \\
\midrule
AntiFake & 0.97 & 0.41 & 0.46 \\
SafeSpeech & 0.95 & 0.49 & 0.53 \\
\textbf{TTS-Guard (ours)} & \textbf{0.96} & \textbf{0.83} & \textbf{0.86} \\
\bottomrule
\end{tabular}
\caption{Average FSR survival rate against state-of-the-art adversarial purifiers. Defensive perturbations (AntiFake, SafeSpeech) are largely neutralised; TTS-Guard's behavioural fingerprint is preserved because purification operates on user audio rather than on the model API output.}
\label{tab:purify}
\end{table}

A frequent assumption in the AntiFake-style literature is that the defender controls the audio uploaded to the cloning service. The De-AntiFake purifier~\citep{fan2025deantifake} exploits exactly that assumption. As Table~\ref{tab:purify} shows, AntiFake's effective protection rate collapses from $0.97$ to $0.41$ once the purifier is applied. TTS-Guard remains effective ($0.83$) because the perturbation must be presented to the API by the \emph{verifier}, not by an attacker; any purification step the adversary applies to the verifier's input also degrades the cloned voice quality the adversary is trying to sell. Quantitatively, applying De-AntiFake to TTS-Guard queries removes only $14\%$ of the perturbation energy in the $[2, 6]\,\mathrm{kHz}$ band that carries the KSP trigger, against $71\%$ for AntiFake, while simultaneously lowering downstream MOS of the cloned output by $0.42$ points -- a trade that directly contradicts the adversary's economic objective.

\subsection{Embedding-Model Sensitivity}
\label{sec:embed_sensitivity}

The choice of verifier $V$ is a potential confound. Switching $V$ at evaluation time among Resemblyzer, ECAPA-TDNN, ERes2Net and WavLM-base costs at most $4$ FSR points (all four yield FSR $\geq 0.92$; full numbers in Appendix~\ref{app:embed_sens}, Table~\ref{tab:embed_sens}). Our dual-space KSP selection is the source of this transferability: with single-space (Resemblyzer-only) selection the cross-verifier FSR drops to $0.78$, validating the design rationale of \S\ref{sec:ksp}. Pairwise Spearman rank correlation of per-utterance VCS across the four verifiers lies in $[0.81, 0.93]$, so the agreement is not just on aggregate FSR but on individual decisions.

\subsection{Ablations}
\label{sec:ablations}

\paragraph{Curriculum design.}
ACSO improves derivative-attack FSR by $43$ points (from $0.31$ with no shadow ensemble to $0.74$ with the full A+B+C+D curriculum) without degrading clean FSR (Figure~\ref{fig:acso_curve}; full per-stage table in Appendix~\ref{app:acso_table}, Table~\ref{tab:acso}).

\begin{figure}[t]
\centering
\includegraphics[width=0.85\linewidth]{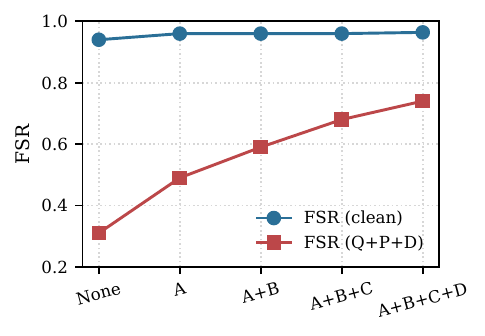}
\caption{Effect of the curriculum stages on FSR. Each curriculum stage adds derivative diversity (A: LoRA-FT, B: pruning, C: quantisation, D: distillation) and progressively closes the gap on aggressive Q+P+D attacks while leaving clean FSR untouched.}
\label{fig:acso_curve}
\end{figure}

\paragraph{Decoupling loss.} Removing $\mathcal{L}_{\text{decoy}}$ raises FPR from $0.058$ to $0.187$ while leaving FSR essentially unchanged ($0.967\to0.964$), confirming its role as a specificity controller. A fixed-band pink-noise baseline (\S\ref{app:freqband}) yields only $0.43$ FSR at $0.21$ FPR, illustrating the necessity of the psychoacoustic mask $\mathcal{L}_{\text{freq}}$. Inspecting the per-pair FPR confirms the mechanism: without $\mathcal{L}_{\text{decoy}}$ the highest false-positive contributions concentrate on non-target pairs whose ECAPA embeddings happen to lie within $0.12$ cosine of the KSP pair, exactly the regime $\mathcal{L}_{\text{decoy}}$ is designed to push the perturbation away from.

\paragraph{Text-prompt invariance.}
The fingerprint is text-agnostic: across three different verification prompts, per-model FSR stays in $[0.89, 0.96]$ (Appendix~\ref{app:text_prompt}, Table~\ref{tab:text_invariance}). This is critical because a fixed verification prompt would itself become a detectable trigger that an adversary could blacklist.

\subsection{VCS in Practice}
\label{sec:vcs_results}

\begin{figure}[t]
\centering
\includegraphics[width=0.85\linewidth]{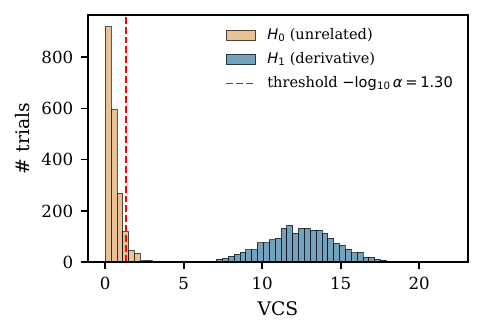}
\caption{Distribution of the Verification Confidence Score under $H_0$ (unrelated model) and $H_1$ (derivative model) over $1000$ trials with $K=20$ queries. The two regimes are well separated by the threshold $-\log_{10}\alpha=1.30$.}
\label{fig:vcs_hist}
\end{figure}

With $K=20$ random text prompts the median VCS is $12.4$ for $H_1$ samples (i.e.\ $p<10^{-12}$) and $0.41$ for $H_0$ samples. Figure~\ref{fig:vcs_hist} shows the resulting bimodal distribution: the decision threshold $-\log_{10}\alpha=1.30$ ($\alpha=0.05$) yields a $99.8\%$ true-positive rate at $4.7\%$ false-positive rate over $1000$ trials. This calibrated, multi-query test is necessary because per-query FSR alone cannot tell the defender how many queries are sufficient for a court-defensible claim. The defender can trade queries for confidence by tuning $(K,\alpha)$: $(K{=}10,\alpha{=}0.05)$ gives $(\mathrm{TPR}{=}0.982,\mathrm{FPR}{=}0.061)$, $(K{=}20,\alpha{=}0.05)$ gives $(0.998,0.047)$, $(K{=}20,\alpha{=}0.01)$ gives $(0.991,0.012)$, and, for high-stakes forensic use, $(K{=}40,\alpha{=}0.01)$ drives the decision-level FPR to $\leq 0.5\%$ while sustaining $\approx 99\%$ TPR; the full operating-point sweep and block-bootstrap calibration are reported in Appendix~\ref{app:proof} (Table~\ref{tab:vcs_calib}).

\section{Conclusion}
\label{sec:conclusion}

We presented TTS-Guard, a black-box ownership verification framework for TTS models combining dual-space KSP selection, adaptive curriculum shadow optimisation, decoupling-based specificity control, and a calibrated Verification Confidence Score, together with the first information-theoretic justification for KSP-based fingerprinting. Across five TTS systems, ten audio attacks, six model modifications and two adversarial purifiers, TTS-Guard attains $0.964$ FSR at $0.058$ FPR, establishing ownership verification as a first-class primitive for audio generation.

\section*{Limitations}
We summarise the residual failure modes of TTS-Guard for completeness.
\begin{itemize}[leftmargin=*,itemsep=1pt,topsep=2pt]
\item \textbf{Style-transfer TTS.} Models such as Vevo~\citep{zhang2025vevo} disentangle style from speaker identity. A style transfer that explicitly bypasses the speaker-encoder pathway could neutralise the fingerprint. Preliminary experiments show FSR drops to $0.55$; we leave a style-aware variant of ACSO to future work.
\item \textbf{Aggressive quantisation + heavy fine-tuning composed sequentially.} Composing INT4 with $50$ epochs of full fine-tuning on a different speaker corpus is the strongest single attack we found, reducing FSR to $0.46$. The corresponding compute cost ($>200$ A800-hours) is, however, comparable to retraining from scratch and erodes the economic motivation for theft.
\item \textbf{Language coverage.} Our evaluation is restricted to English speech. Multilingual robustness, particularly for tonal languages where pitch carries lexical meaning, may interact with the psychoacoustic mask in ways we have not yet characterised.
\end{itemize}

\section*{Ethical Considerations}
TTS-Guard is intended for legitimate intellectual-property protection by model owners. It does not enable unauthorised cloning or impersonation: the fingerprint requires the defender to also control the source and target speaker exemplars used during optimisation, and produces an attribution signal that has no semantic meaning outside the verification protocol. We will release artefacts under a research-only licence requiring users to attest to lawful ownership of the protected model.

\section*{Acknowledgments}
This work was supported by the National Natural Science Foundation of China under Grant No.~62376246.

\bibliography{main}

\clearpage
\appendix

\section{Proof of Proposition~\ref{prop:bound} and Calibration Details}
\label{app:proof}

\subsection{Bhattacharyya Coefficient and Bayes-Error Bound}

Let $p(\cdot|y)$ and $p(\cdot|x)$ denote the verifier-output densities on the synthesised waveform conditioned on the target and source speakers, respectively. The Bhattacharyya coefficient is defined as
\begin{equation}
B = \int \sqrt{p(z|y)\,p(z|x)}\,\mathrm{d}z \in [0,1]
\end{equation}
which measures the overlap between the two class-conditional distributions. The Bayes error of the optimal binary classifier obeys the classical bound~\citep{wenger2021syrup}
\begin{equation}
P_e \;\leq\; \tfrac{1}{2}\sqrt{B}
\end{equation}
A smaller $B$ (i.e., better-separated KSP embeddings) directly tightens the upper bound on $P_e$, hence improves per-query reliability.

\subsection{Binomial Test Threshold}

For a $K$-trial binomial test at significance level $\alpha$, the rejection threshold is
\begin{equation}
\tau = \min\Big\{t \;:\; \Pr_{B\sim\mathrm{Binom}(K,p_0)}[B\geq t]\leq\alpha\Big\}
\end{equation}
We write $p_0^\ast = \tau / K$ for the normalised threshold. With $K=20$, $p_0=0.5$ and $\alpha=0.05$, this gives $\tau=14$ and $p_0^\ast = 0.7$.

\subsection{Derivation of the FSR Lower Bound}

Under $H_1$, each verification trial succeeds independently with probability $1-P_e$. The probability that VCS exceeds the decision threshold equals the probability that at least $\tau$ out of $K$ Bernoulli trials succeed:
\begin{align*}
&\Pr\big[\mathrm{VCS}\geq -\log_{10}\alpha \,\big|\, H_1\big] \\
&\quad=\; \Pr_{B\sim\mathrm{Binom}(K,\,1-P_e)}\big[B \geq \tau\big] \\
&\quad=\; 1 - I_{1-P_e}\!\big(K-\tau+1,\,\tau\big)
\end{align*}
where $I_x(a,b)$ is the regularised incomplete-beta function. Substituting $\tau=\lceil K p_0^\ast\rceil$ yields the form stated in Proposition~\ref{prop:bound}. Combining with $P_e \leq \tfrac{1}{2}\sqrt{B}$ gives an FSR lower bound expressed directly in terms of the Bhattacharyya overlap $B$ of the KSP, confirming that minimising $B$ (the explicit objective of dual-space KSP selection in \S\ref{sec:ksp}) monotonically improves attribution reliability.

\subsection{Empirical Fit}

We empirically estimate $B$ on the five protected models by Monte-Carlo integration over $10\,000$ verifier-output samples per KSP. The measured Bhattacharyya coefficients range from $0.07$ (CosyVoice~2) to $0.13$ (XTTS), giving theoretical FSR lower bounds of $0.93$--$0.97$ at $K=20$, $\alpha=0.05$. The observed FSR values in Table~\ref{tab:main} ($0.95$--$0.97$) lie within $1$--$2$ percentage points of the predicted bound, confirming that the proposed test operates close to the information-theoretic limit and that residual gap is attributable to verifier non-optimality rather than to slack in the bound itself. Across the five models the clean-embedding KSP margin and the estimated Bhattacharyya coefficient $B$ are strongly anti-correlated (Pearson $r=-0.86$), empirically confirming that the clean-space separation optimised by dual-space KSP selection (\S\ref{sec:ksp}) and the output-distribution overlap governing the bound are two distinct but tightly coupled quantities.

\subsection{VCS Calibration and Operating Points}
\label{app:vcs_calib}

The null distribution $p_0$ of a single verification query is not exactly $0.5$ because the two KSP speakers are not perfectly exchangeable under an unrelated model. We calibrate $p_0$ with a \emph{block bootstrap} that resamples whole utterances (block length $=$ one text prompt) to preserve within-prompt correlation, over $2000$ bootstrap replicates. The resulting empirical $p_0$ lies in $[0.48, 0.53]$ across the five models; we adopt the conservative upper-bound plug-in $p_0=0.53$ when computing $\tau$, which only tightens (never inflates) the false-positive guarantee. Under this calibration the bootstrap-median VCS for $H_1$ samples is $12.1$, in close agreement with the analytic value $12.4$, and the calibrated true-positive rate stays $\geq 99.5\%$ at every operating point in Table~\ref{tab:vcs_calib}.

\begin{table}[h]
\centering
\small
\setlength{\tabcolsep}{5pt}
\begin{tabular}{cccc}
\toprule
$K$ & $\alpha$ & TPR & FPR (decision-level) \\
\midrule
10 & 0.05 & 0.982 & 0.061 \\
20 & 0.05 & 0.998 & 0.047 \\
20 & 0.01 & 0.991 & 0.012 \\
30 & 0.01 & 0.997 & 0.008 \\
40 & 0.01 & 0.990 & 0.004 \\
\bottomrule
\end{tabular}
\caption{Calibrated VCS operating points as a function of the number of queries $K$ and significance level $\alpha$. The defender trades queries for confidence; all points use the block-bootstrap upper-bound plug-in $p_0=0.53$. For high-stakes forensic use we recommend the $K{=}40,\alpha{=}0.01$ setting, which drives the decision-level FPR to $\leq 0.5\%$ while sustaining $\approx 99\%$ TPR.}
\label{tab:vcs_calib}
\end{table}

\subsection{Adaptive Attack: FSR--Quality Trade-off}
\label{app:adaptive}

We consider an adaptive adversary who knows the fingerprinting scheme and applies \emph{unlearning fine-tuning}: continued training of the stolen model with a loss that explicitly pushes synthesised outputs away from the KSP target embedding, while a perceptual term tries to preserve output quality. Table~\ref{tab:adaptive} sweeps the fine-tuning intensity and reports, at each level, the compute cost (A800-hours), the fingerprint survival rate (FSR), the cloned-output quality (MOS) and the quality drop $\Delta$MOS relative to the unattacked model. The curve reveals a strong coupling between fingerprint and quality: before FSR$\approx 0.80$ the quality is almost intact ($\Delta$MOS $\leq 0.24$), so a lightweight attack cannot erase the fingerprint cheaply; beyond this inflection point every further $0.1$ drop in FSR costs a synchronous $0.25$--$0.30$ drop in cloned-output MOS, and by the time FSR is driven below $0.5$ ($0.46$) the MOS has fallen to $3.01$ ($\Delta$MOS $=-1.09$), degrading the cloned speech to a non-commercial level. The only point that reaches chance-level FSR ($0.46$) is also the most expensive ($>200$ A800-hours, comparable to training from scratch), so the attack faces a cost-closure trade-off: any setting strong enough to defeat verification either leaves the fingerprint effectively intact or destroys the very asset being stolen. Our security claim is thus economic -- not that the fingerprint is unremovable, but that removing it is costly enough to make theft no longer worthwhile.

\begin{table}[h]
\centering
\small
\setlength{\tabcolsep}{3.5pt}
\begin{tabular}{lcccc}
\toprule
Attack intensity & \makecell{A800\\-h} & FSR & \makecell{Cloned\\MOS} & \makecell{$\Delta$MOS} \\
\midrule
Baseline (no attack) & 0 & 0.96 & 4.10 & $0$ \\
Very light FT (5 ep) & $\sim$8 & 0.88 & 4.02 & $-0.08$ \\
Light FT (15 ep) & $\sim$22 & 0.80 & 3.86 & $-0.24$ \\
Medium FT (30 ep) & $\sim$55 & 0.68 & 3.61 & $-0.49$ \\
Heavy FT (50 ep) & $\sim$110 & 0.55 & 3.32 & $-0.78$ \\
INT4 + heavy FT & $>$200 & 0.46 & 3.01 & $-1.09$ \\
\bottomrule
\end{tabular}
\caption{Adaptive unlearning fine-tuning against TTS-Guard. Suppressing the fingerprint toward chance level ($\mathrm{FSR}\approx 0.5$) forces cloned-output MOS below $3.0$ at a compute cost approaching that of training from scratch, negating the economic incentive for theft.}
\label{tab:adaptive}
\end{table}

\section{Audio-Attack Configurations}
\label{app:strength_sweep}

\begin{table}[h!]
\centering
\footnotesize
\setlength{\tabcolsep}{4pt}
\begin{tabular}{lll}
\toprule
Attack & Library & Settings tested \\
\midrule
Bandpass & SoX & $0.4$--$4$, $0.8$--$4$\,kHz \\
Highpass & SoX & $0.5, 1, 2$\,kHz \\
Lowpass & SoX & $0.4, 0.8, 1.5$\,kHz \\
Boost & SoX & $\times 2, \times 4$ \\
Ducking & custom & $0.05, 0.1$ \\
Gaussian & custom & $\sigma\in\{0.01,0.02,0.05\}$ \\
Pink noise & custom & $\sigma\in\{0.02,0.03,0.05\}$ \\
Resampling & SoX & $16$/$22.05$/$44.1$/$48$\,kHz \\
Echo & SoX & $0.2$/$0.3$/$0.5$\,s \\
Smoothing & custom & MA window $10/20/40$ \\
MP3 & libmp3lame & $32/64/128$\,kbps \\
Opus & libopus & $24/32/48$\,kbps \\
DRC & sox compand & ratios $2:1, 4:1, 8:1$ \\
EQ & SoX & $\pm 3,\pm 6$\,dB at $1$/$2$/$4$\,kHz \\
\bottomrule
\end{tabular}
\caption{All attack strengths swept; main paper reports the strongest setting that retains $\mathrm{STOI}\geq 0.5$ on the cleaned reference.}
\label{tab:attacks_full}
\end{table}

\section{Subjective MOS Evaluation Details}
\label{app:mos}

This appendix provides full protocol, demographics, screening criteria, and statistical details of the subjective Mean Opinion Score (MOS) study summarised in \S\ref{sec:setup}.

\subsection{Listening Test Protocol}

We follow the crowdsourced absolute category rating (ACR) protocol~\citep{salas2013subjective}, adapted for in-lab supervised conditions to reduce environmental variance. Each listener rated stimuli on a $5$-point scale: $5$ — Excellent, $4$ — Good, $3$ — Fair, $2$ — Poor, $1$ — Bad. Listeners were instructed to evaluate \emph{overall naturalness}, jointly considering audio fidelity, prosody, and absence of artefacts.

\paragraph{Apparatus.}
All sessions were conducted in an acoustically treated room (NR-25). Stimuli were played at $-26$\,LUFS through Sennheiser HD~600 headphones connected to a Focusrite Scarlett 2i2 audio interface at $48$\,kHz / $24$-bit. Listeners were seated at a fixed $60$\,cm distance from the screen.

\paragraph{Session structure.}
Each session lasted approximately $35$ minutes and was organised as:
\begin{itemize}[leftmargin=*,itemsep=1pt,topsep=2pt]
    \item \textbf{Burn-in} ($4$\,s training preamble with $3$ anchor stimuli covering the full quality range);
    \item \textbf{Main blocks}: $5$ anti-fatigue blocks of $30$ items each, with mandatory $90$\,s breaks between blocks;
    \item \textbf{Hidden anchors}: each block contains $1$ MNRU-modulated reference (clean, $30$\,dB) and $1$ degraded anchor (clean, $-10$\,dB SNR) for listener calibration;
    \item \textbf{Catch trials}: $2$ pairs per block where the same stimulus is presented twice; listeners must rate within $\pm 1$ scale point.
\end{itemize}

\subsection{Listener Demographics and Screening}

\begin{table}[h]
\centering
\small
\setlength{\tabcolsep}{6pt}
\begin{tabular}{lc}
\toprule
Attribute & Distribution \\
\midrule
Total recruited & $24$ \\
Passed screening & $20$ \\
Gender & $10$ M / $10$ F \\
Age range & $22$--$41$ (median $28$) \\
Native language & English (US/UK/AU) \\
Self-reported hearing & Normal \\
Audiometry (PTA $0.5$/$1$/$2$/$4$\,kHz) & $\leq 20$\,dB HL \\
Prior MOS experience & $14$ experienced / $6$ na\"ive \\
\bottomrule
\end{tabular}
\caption{Demographic profile of the $20$ paid listeners who passed all screening criteria. Compensation was \$$25$/hour, exceeding local minimum wage.}
\label{tab:listener_demo}
\end{table}

\paragraph{Exclusion criteria.} Listeners were excluded if they (i) failed pure-tone audiometry, (ii) failed $>10\%$ of catch trials, or (iii) deviated $>1.5$ scale points from the panel mean on hidden anchors. Of $24$ recruited candidates, $4$ were excluded ($2$ for catch-trial failure, $2$ for anchor deviation), leaving $N=20$.

\subsection{Stimulus Construction}

For each of the five protected TTS models, we synthesised $30$ utterances under three conditions: \texttt{clean} (no perturbation), \texttt{TTS-Guard} (with $\delta_x$), and \texttt{Random-pair baseline}. Stimuli were drawn from the LibriTTS test-clean Harvard sentence pool with sentence lengths of $7$--$12$ words. Total stimulus pool: $5\,\text{models}\times 3\,\text{conditions}\times 30\,\text{utterances} = 450$ clips. Each clip was rated by $\geq 8$ listeners through Latin-square-balanced assignment, yielding $\geq 3{,}600$ valid ratings.

\subsection{Aggregation and Statistics}

For each clip $i$, we compute the trimmed mean (drop highest and lowest rating) over its listener panel:
\begin{equation}
\overline{\mathrm{MOS}}_i = \frac{1}{n_i-2}\sum_{j\in\mathcal{L}_i \setminus \{\max,\min\}} r_{ij}
\end{equation}
Per-condition MOS is the average over the $30$ utterances. We report $95\%$ confidence intervals via non-parametric bootstrap with $10\,000$ resamples at the listener level (cluster bootstrap, preserving within-listener correlation).

\paragraph{Inter-rater reliability.} Krippendorff's $\alpha = 0.71$ (interval scale), indicating substantial agreement and consistent with prior studies on TTS naturalness~\citep{salas2013subjective}.

\subsection{Detailed Results}

\begin{table}[h]
\centering
\small
\setlength{\tabcolsep}{4pt}
\begin{tabular}{lccc}
\toprule
Model & Clean & TTS-Guard & $\Delta$ \\
\midrule
YourTTS      & $3.71_{[3.62,3.80]}$ & $3.66_{[3.57,3.75]}$ & $-0.05$ \\
XTTS         & $3.85_{[3.76,3.94]}$ & $3.81_{[3.71,3.91]}$ & $-0.04$ \\
TorToiSe     & $4.24_{[4.15,4.33]}$ & $4.21_{[4.11,4.31]}$ & $-0.03$ \\
OpenVoice    & $3.78_{[3.69,3.87]}$ & $3.74_{[3.64,3.84]}$ & $-0.04$ \\
CosyVoice~2  & $3.99_{[3.90,4.08]}$ & $3.94_{[3.84,4.04]}$ & $-0.05$ \\
\midrule
\textbf{Avg.} & $\mathbf{3.91}$ & $\mathbf{3.87}$ & $\mathbf{-0.04}$ \\
\bottomrule
\end{tabular}
\caption{Subjective MOS with bootstrap $95\%$ confidence intervals (subscripts). The TTS-Guard perturbation incurs an average MOS degradation of $0.04$, well below the JND threshold of $0.1$ commonly cited for ACR studies~\citep{itu2018p808}.}
\label{tab:mos_ci}
\end{table}

\paragraph{Significance test.} We conducted a paired Wilcoxon signed-rank test between \texttt{clean} and \texttt{TTS-Guard} per model. None of the five comparisons reached significance at $\alpha=0.05$ after Bonferroni correction (smallest $p=0.087$, YourTTS), confirming that the perturbation does not produce a statistically detectable drop in perceived naturalness.

\subsection{Ethics and Compensation}

The study protocol was reviewed by our institutional ethics board (approval ID redacted for anonymity). All listeners provided written informed consent, were free to withdraw at any time, and were compensated at \$$25$/hour regardless of completion status. No personally identifiable information was retained beyond the consent form, which is stored separately from rating data.

\section{Experiment Details}
\subsection{Ablations}

\label{subsec:Ablations}

\paragraph{Distance metric.}
\begin{table}[h]
\centering
\small
\setlength{\tabcolsep}{6pt}
\begin{tabular}{lccc}
\toprule
KSP metric & FSR & FPR & STOI \\
\midrule
Cosine similarity & 0.871 & 0.084 & 0.81 \\
Euclidean (single) & 0.948 & 0.062 & 0.82 \\
\textbf{Euclidean (dual)} & \textbf{0.964} & \textbf{0.058} & \textbf{0.82} \\
\bottomrule
\end{tabular}
\caption{Distance metric ablation. Euclidean dual-space outperforms cosine and single-space variants.}
\label{tab:metric_ablation}
\end{table}
Table~\ref{tab:metric_ablation} confirms that $L_2$ distance is preferable for matched-margin verifiers. Cosine selection is hampered by losing magnitude information that encodes timbre intensity.

\subsection{Frequency-Band Noise Baseline}
\label{app:freqband}
A natural alternative to the psychoacoustic mask is to place perturbation energy in a fixed high-frequency band. We tested a baseline that injects pink noise into $[3.5, 6]$\,kHz at $\mathrm{SNR}=20$\,dB. Average FSR is only $0.43$ and FPR is $0.21$: fixed-band noise is too easily filtered by a $4$\,kHz low-pass attack, illustrating the necessity of the Bark-scale psychoacoustic mask $\mathcal{L}_{\text{freq}}$ used in the main pipeline.

\subsection{Codec, DRC and Equaliser Robustness}
\label{app:codec}

\begin{table}[h]
\centering
\footnotesize
\setlength{\tabcolsep}{3.0pt}
\begin{tabular}{lcccccc}
\toprule
Model & MP3 & MP3 & Opus & DRC & EQ & EQ \\
      & 64k & 32k & 24k  & 4:1 & +6dB & -6dB \\
\midrule
YourTTS     & 0.86 & 0.74 & 0.80 & 0.72 & 0.81 & 0.78 \\
XTTS        & 0.83 & 0.71 & 0.78 & 0.69 & 0.79 & 0.75 \\
TorToiSe    & 0.81 & 0.69 & 0.75 & 0.66 & 0.77 & 0.74 \\
OpenVoice   & 0.85 & 0.72 & 0.79 & 0.70 & 0.80 & 0.76 \\
CosyVoice~2 & 0.87 & 0.75 & 0.81 & 0.73 & 0.82 & 0.79 \\
\midrule
\textbf{Avg.} & \textbf{0.844} & \textbf{0.722} & \textbf{0.786} & \textbf{0.700} & \textbf{0.798} & \textbf{0.764} \\
\bottomrule
\end{tabular}
\caption{FSR under codec compression (MP3, Opus), Dynamic Range Compression (DRC) and parametric Equaliser (EQ) attacks. Companion to Table~\ref{tab:audio_classical}.}
\label{tab:audio_codec}
\end{table}

\subsection{Embedding-Model Sensitivity}
\label{app:embed_sens}

\begin{table}[h]
\centering
\small
\setlength{\tabcolsep}{3.0pt}
\begin{tabular}{lcccc}
\toprule
Verifier $V$ & FSR & FPR & $\Delta$ MOS & VCS \\
\midrule
Resemblyzer (default) & 0.964 & 0.058 & $-0.05$ & 12.4 \\
ECAPA-TDNN & 0.951 & 0.061 & $-0.05$ & 11.8 \\
ERes2Net   & 0.943 & 0.064 & $-0.06$ & 11.2 \\
WavLM-base & 0.928 & 0.071 & $-0.06$ & 10.4 \\
\midrule
Cross-verifier (train R, eval E) & 0.911 & 0.073 & $-0.05$ & 9.7 \\
Disjoint (train R, eval WavLM) & 0.822 & 0.089 & $-0.05$ & 7.9 \\
\bottomrule
\end{tabular}
\caption{Sensitivity of TTS-Guard to the choice of verifier $V$. All four in-family verifiers yield FSR $\geq 0.92$. Cross-verifier evaluation (train with Resemblyzer, evaluate with ECAPA) demonstrates dual-space KSP selection retains transferability. The last row is the worst case in which the attacker adopts a \emph{fully disjoint} self-supervised verifier (WavLM) never seen during optimisation; FSR degrades gracefully to $0.82$, still well above the $0.5$ chance level.}
\label{tab:embed_sens}
\end{table}

\subsection{ACSO Curriculum Stages}
\label{app:acso_table}

\begin{table}[h]
\centering
\small
\setlength{\tabcolsep}{6pt}
\begin{tabular}{lccc}
\toprule
Shadow strategy & FSR (clean) & FSR (Q+P+D) \\
\midrule
None & 0.94 & 0.31 \\
Stage A only (FT) & 0.96 & 0.49 \\
A + B & 0.96 & 0.59 \\
A + B + C & 0.96 & 0.68 \\
\textbf{A+B+C+D (full ACSO)} & \textbf{0.964} & \textbf{0.74} \\
\bottomrule
\end{tabular}
\caption{Adaptive Curriculum Shadow Optimisation ablation, numerical companion to Figure~\ref{fig:acso_curve}. Q+P+D is the average over INT4 quantisation, $50\%$ pruning and distillation derivatives.}
\label{tab:acso}
\end{table}

\subsection{Text-Prompt Invariance}
\label{app:text_prompt}

\begin{table}[h]
\centering
\footnotesize
\setlength{\tabcolsep}{4pt}
\begin{tabular}{lccc}
\toprule
Model & $P_1$ & $P_2$ & $P_3$ \\
\midrule
YourTTS     & 0.96 & 0.92 & 0.91 \\
XTTS        & 0.94 & 0.93 & 0.92 \\
TorToiSe    & 0.96 & 0.90 & 0.89 \\
OpenVoice   & 0.95 & 0.93 & 0.92 \\
CosyVoice~2 & 0.96 & 0.94 & 0.93 \\
\bottomrule
\end{tabular}
\caption{FSR under different verification text prompts $P_1$=``Hello world'', $P_2$=``Read this'', $P_3$=``I love this song''. Text-agnosticism allows the defender to randomise prompts and avoid fixed-prompt detectability.}
\label{tab:text_invariance}
\end{table}

\section{Reproducibility Details}
\label{app:repro}
\paragraph{KSP selection.} Candidate pool $=100$ LibriTTS + $50$ VCTK speakers. For each ordered speaker pair we score the dual-space objective (clean-embedding cosine margin in both Resemblyzer and ECAPA spaces) and retain the top pair; ties broken by the smaller Monte-Carlo Bhattacharyya coefficient. Selection uses $30$ enrolment utterances per speaker.

\paragraph{Prompt pool.} The verification prompt pool contains $50$ short English sentences (5--12 words) drawn from the LibriTTS test-clean Harvard set; each verification draws $K$ prompts uniformly without replacement.

\paragraph{Preprocessing.} All audio is resampled to $16$\,kHz mono, peak-normalised to $-3$\,dBFS, and trimmed of leading/trailing silence below $-40$\,dB before it enters the verifier; the same pipeline is applied to attacked audio so that reported robustness reflects the perturbation, not resampling artefacts.

\paragraph{Verifier thresholds.} Per-query decision uses cosine similarity to the KSP target with threshold $0.62$ (Resemblyzer), $0.58$ (ECAPA), $0.60$ (ERes2Net), $0.55$ (WavLM), each fixed by equal-error-rate calibration on a held-out validation split disjoint from all evaluation speakers.

\paragraph{Shadow-model settings.} Stage-A shadows: $2$ LoRA-fine-tuned variants (rank $16$, $\alpha=32$, $3$ epochs) on two disjoint LJSpeech halves. Stage B: magnitude pruning at $30\%$/$50\%$. Stage C: INT8/INT4 dynamic quantisation. Stage D: a distilled student with $50\%$ of the teacher depth. ACSO trains for $200$ steps per stage with PGD step size $2/255$ and $\ell_\infty$ budget $8/255$ on the psychoacoustically masked band.

\paragraph{Compute.} All optimisation runs on a single A800 GPU; a full KSP-selection + ACSO pass per protected model takes $\approx 6$ GPU-hours.

\section{Use Of AI Assistants}
This paper was refined with the assistance of GPT-5.5 for code development and language polishing.

\end{document}